# Draping of the Interstellar Magnetic Field over the Heliopause - A Passive Field Model

Philip A. Isenberg, Terry G. Forbes & Eberhard Möbius

Space Science Center and Department of Physics, University of New Hampshire, Durham, NH 03824, USA

**ABSTRACT**

As the local interstellar plasma flows past our heliosphere, it is slowed and deflected around the magnetic obstacle of the heliopause. The interstellar magnetic field, frozen into this plasma, then becomes draped around the heliopause in a characteristic manner. We derive the analytical solution for this draped magnetic field in the limit of weak field intensity, assuming an ideal potential flow around the heliopause, which we model as a Rankine half-body. We compare the structure of the model magnetic field with observed properties of the IBEX ribbon and with *in situ* observations at the Voyager 1 spacecraft. We find reasonable qualitative agreement, given the idealizations of the model. This agreement lends support to the secondary ENA model of the IBEX ribbon and to the interpretation that Voyager 1 has crossed the heliopause. We also predict that the magnetic field measured by Voyager 2 after it crosses the heliopause will not be significantly rotated away from the direction of the undisturbed interstellar field.

## 1. Introduction

Over the last several decades, we have obtained considerable useful information on the properties of the local interstellar medium (LISM) just beyond the heliopause boundary of our heliosphere. The overall speed and direction of the interstellar fluid flow relative to the Sun has been generally known for several decades from observations of the neutral helium atoms streaming into the heliosphere (Gloeckler et al. 2004; Möbius et al. 2004; Witte 2004), though specific values are still under some discussion (Möbius et al. 2012; Bzowski et al. 2012; Frisch et al. 2013, 2015; Lallement & Bertaux 2014; McComas et al. 2015; Wood et al. 2015; Leonard et al. 2015).

More recently, properties of the interstellar magnetic field (ISMF) have become available through combinations of indirect inferences from models and direct observation. Estimates of the tilt angle between the flow direction and the magnetic field were first suggested on the basis of the hydrogen deflection plane (Lallement et al. 2005; Izmodenov et al. 2005; Lallement et al. 2010) and on the observed asymmetry of the termination shock crossings by the two Voyager spacecraft (Opher et al. 2006; Opher et al. 2007; Opher et al. 2009; Pogorelov et al. 2007; Pogorelov et al. 2008; Pogorelov et al. 2009; Ratkiewicz & Grygorczuk 2008).

Subsequently, the discovery by the Interstellar Boundary Explorer (IBEX) of a ribbon of enhanced energetic neutral atom (ENA) flux (McComas et al. 2009; Funsten et al. 2009; Fuselier et al. 2009) has lead to more detailed estimates. It is widely understood that the IBEX ribbon indicates the location where the ISMF is perpendicular to the heliospheric radial direction, $\mathbf{B} \cdot \hat{\mathbf{r}} = 0$ (Schwadron et al. 2009; Ratkiewicz et al. 2012; McComas et al. 2014). In particular, we focus here on the various versions of the 'secondary ENA scenario' (McComas et al. 2009; Heerikhuisen et al. 2010; McComas et al. 2014). This scenario proposes that the ribbon results from a multi-step process which begins with the charge-exchange ionization of slowly moving interstellar neutral hydrogen in the solar wind and inner heliosheath, leading to primary ENAs flowing anti-sunward into the local interstellar medium (LISM). These atoms are then charge-exchange ionized by the neutral component of the LISM and the pickup protons that are produced gyrate around the local magnetic field. These pickup protons are somehow confined either in pitch angle (Heerikhuisen et al. 2010; Florinski et al. 2010; Chalov et al. 2010; Gamayunov et al. 2010; Möbius et al. 2013) or in space (Schwadron & McComas 2013a; Isenberg 2014) until they are again neutralized by charge exchange. The portion of these secondary ENAs with the appropriate gyrophase at neutralization will be observed by IBEX from the region where $\mathbf{B} \cdot \hat{\mathbf{r}} = 0$.

Under this scenario, if the ISMF were uniform and unaffected by the presence of the heliosphere, the IBEX ribbon would appear as a great circle in the sky and the center of this circle would designate the direction of the field (Schwadron et al. 2009; Funsten et al. 2009; Funsten et al. 2013; Grygorczuk et al. 2011). The observed ribbon does appear roughly as a circle in the sky with a center near ecliptic latitude 40° and ecliptic longitude 219°, but the opening angle of the circle is less than 90° (Funsten et al. 2013). This property can be understood under the secondary ENA scenario if the ribbon ENAs are primarily emitted from regions where the ISMF is deflected by the relative flow of the heliosphere through the LISM. This interpretation is further supported by the fact that the opening angle and the apparent center of the circle shift with ENA energy as would be expected if the more energetic ENAs come from more distant positions. The indicated orientation of the ISMF is also consistent with the recently measured anisotropy of TeV cosmic rays (Schwadron et al. 2014; Zhang et al. 2014).

The *in situ* measurements of the magnetic field at the Voyager 1 spacecraft, now beyond 130 AU from the Sun, yield an independent determination of the field strength and orientation just beyond the heliopause (Burlaga & Ness 2014a, b). These fields have been advected around the obstacle presented by the heliosphere as it moves through the interstellar plasma. A model of the draped ISMF which is consistent with both the IBEX ribbon determination and the Voyager measurements would provide supporting evidence for the secondary ENA scenario and yield important information on the large-scale structure of the heliopause.

At present, studies of the heliospheric structure in the presence of a magnetized LISM flow rely primarily on detailed and complicated MHD simulations of the global interactions between heliospheric and interstellar partially-ionized media (Pogorelov et al. 2011; Heerikhuisen & Pogorelov 2011; Heerikhuisen et al. 2014; Strumik et al. 2011; Grygorczuk et al. 2011; Grygorczuk et al. 2014; Ben-Jaffel et al. 2013; Opher & Drake 2013). These involved studies have been motivated by the understanding that the magnetic pressure of the ISMF and the dynamical pressure of the LISM flow appear to be comparable, and a rigorous model needs to include the nonlinear combination of both pressures. We do not disagree with this assessment, but we will show in this paper that useful information may also be obtained from simpler, linear calculations of fluid flow around a predefined obstacle, in the limit of vanishing magnetic forces.

Ultimately, by considering a variety of idealized shapes for the heliospheric obstacle and comparing the results of these calculations with the available observations, we hope to gain a qualitative understanding of the heliospheric interaction with the interstellar medium without resorting to massive simulations. The model presented here is intended as a first step in such a program.

A similar progression in the opposite limit of vanishing dynamic force of the interstellar plasma flow can be found by considering several other analytical results. In a part of his influential paper, Parker (1961) derived the interstellar magnetic field surrounding a spherical outflow from a point source, representing a stellar wind in a very strong magnetic field. Whang (2010) and Schwadron et al. (2014) have extended that picture by imposing heliopause shapes with a nose and a tail, as expected to be formed by the flow of the LISM around the heliosphere. However, these papers did not actually include a flow.

In this paper, we will describe the linear problem of a passive magnetic field, frozen into a potential plasma flow around a rigid, predefined obstacle. In this first step, we will take the heliospheric obstacle to be a Rankine half-body, with its symmetry axis aligned with the flow. This case affords the additional advantage of an analytic solution to the problem, so the entire structure of the field and flow may be explored. We will see that this solution has some interesting correspondence with the observations, but that some details indicate the need for a more complicated shape for the heliopause.

Subsequent to the submission of this paper, we learned of a work addressing the same mathematical problem (Röken et al. 2015) which had been submitted less than a month earlier than ours. Despite many similarities, the mathematical treatment here is somewhat different, employing a different coordinate system and therefore obtaining different, though compatible, expressions for the solution. We note as well that the applications discussed in each paper are quite independent, with Röken et al. comparing the solutions to those of MHD simulations and our paper addressing the observational implications for the ISMF upstream of our heliosphere. Thus, distinct contributions to our understanding of the LISM are provided in each paper.

In the next section, we describe the ideal problem of a passive frozen-in magnetic field carried around a Rankine half-body. In Section 3, we present the mathematical analysis and detailed solution. We then apply this solution to the case of our heliosphere, discussing the implications for the measurements of the IBEX ribbon in Section 4, and the Voyager measurements in Section 5. Section 6 contains a summary and our conclusions.

## 2. The general passive-field problem

Consider the idealized problem of steady potential fluid flow past an obstacle. We take the flow at upstream infinity to be constant and uniform. The velocity field is incompressible,

∇ • **V** = 0, and is derived from a scalar potential, **V** = ∇$\phi$, so that $\nabla^2 \phi = 0$. Taking the boundary of the obstacle to be frictionless, solutions of this Laplace equation can be obtained in closed form for a wide variety of obstacle shapes, so the velocity field is a known function of position.

When this fluid is perfectly conducting, a passive magnetic field, **B**, will satisfy the frozen-in condition, ∇ × (**V** × **B**) = 0. We also take the magnetic field at upstream infinity to be constant and uniform. Since **V** is known, we are left with a set of coupled linear differential equations for **B**, which can be solved by standard analytical or numerical methods.

If the flow around the obstacle is axisymmetric, or otherwise effectively two-dimensional, it can be described by a scalar stream function and an analytical solution for **B** may then be obtained. Solutions for the passive magnetic field carried around a spherical obstacle exist in the literature (Bernikov & Semenov 1979; Chacko & Hassam 1997; Dursi & Pfrommer 2008; Romanelli et al. 2014). The passive magnetic field around an axisymmetric, but numerically-determined, heliopause has also been studied (Mitchell et al. 2008). For the present context of the LISM flow past the heliosphere, we consider the case of a Rankine half-body (Batchelor 1967), shown in Figure 1.

The Rankine half-body is formed by the superposition of two basic flows: a uniform flow and flow from a three-dimensional point source. In spherical coordinates, where the point source is at the origin and the polar axis is directed opposite to the uniform flow, the velocity field can be written as

$$\mathbf{V} = \left(\frac{1}{r^2} - \cos\theta\right)\hat{\mathbf{r}} + \sin\theta\,\hat{\boldsymbol{\theta}}. \tag{1}$$

Here, we have normalized the respective flows so that the stagnation point along the polar axis appears at $r = 1$, and in the remainder of this paper all distances will be scaled to that length. In the context of the flow around the heliosphere, this length corresponds to the distance between the Sun and the nose of the heliopause.

In this potential velocity field, the flows from the two sources do not interpenetrate. This can be seen from the stream function, $\Psi$, which specifies a fluid streamline as $\Psi$ = constant. The stream function for this flow is obtained from the superposition of the individual stream functions, so that

$$\Psi = 1 - \cos\theta - \frac{r^2}{2}\sin^2\theta. \qquad (2)$$

With this definition, the stagnation point of the flow appears at $\Psi = 0$. Thus, the surface obtained from this condition, $r^2 \sin^2\theta = 2(1 - \cos\theta)$, separates the two fluids and acts as an obstacle in the external flow. The obstacle shape, identified here as the heliopause in the interstellar flow and shown in red in Figure 1, is more easily recognized using the "impact parameter", $p = r \sin\theta$, giving a heliopause shape as

$$p_{HP} = \sqrt{2(1 - \cos\theta)}. \qquad (3)$$

The properties of the external flow do not depend on conditions inside this model heliopause, so we treat this surface as a rigid boundary in the external flow without further concern of the internal problem.

We note that the Rankine half-body flow is identical to the fluid flow derived by Parker (1961) in his treatment of the heliospheric shape in the limit of a weak ISMF. Parker did not proceed to solve the problem of the resulting passive magnetic field, which we address in the next section.

## 3. Mathematical Solution

As stated above, the passive magnetic field frozen into a known plasma flow is the solution of a set of linear differential equations given by the frozen-flux condition, $\nabla \times (\mathbf{V} \times \mathbf{B}) = 0$, and a specification of the field at upstream infinity. The basic procedure for solving this system in the case of flow around a spherical obstacle is outlined in Bernikov & Semenov (1979) and presented more fully in the appendices of Dursi & Pfrommer (2008). Those papers obtained solutions which, though valid and usable, formally diverge at large distances from the obstacle. In this paper, we will proceed in a different manner, using geometrical arguments and the Euler potentials (e.g. Stern 1966, 1970) to describe the field. We will see that this method results in expressions for the magnetic field which are well behaved everywhere away from the obstacle surface. In the Appendix, we apply this procedure to the spherical obstacle case and present the equivalent expressions for that problem.

For our case of interstellar flow around a Rankine half-body heliosphere, we take the field at upstream infinity to be uniform and constant. The solution for the total field will be the linear superposition of the solution for a field which starts out at infinity parallel to the flow (the longitudinal component, $B_\ell$), and the solution for a field which at infinity is perpendicular to the flow (the transverse component, $B_t$),

$$\mathbf{B} = \mathbf{B}_\ell + \mathbf{B}_t. \tag{4}$$

In the heliocentric spherical coordinate system with the polar axis pointed into the fluid flow, and designating the field intensity at infinity as $B_o$, we have the boundary conditions,

$$\mathbf{B}_{\ell o} = B_o \cos\theta_{BV}(-\cos\theta\, \hat{\mathbf{r}} + \sin\theta\, \hat{\boldsymbol{\theta}})$$

$$\mathbf{B}_{to} = B_o \sin\theta_{BV}\left[\sin\theta \cos(\phi-\phi_o)\hat{\mathbf{r}} + \cos\theta \cos(\phi-\phi_o)\hat{\boldsymbol{\theta}} - \sin(\phi-\phi_o)\hat{\boldsymbol{\phi}}\right], \tag{5}$$

where $\theta_{BV}$ is the angle between the uniform flow and the uniform field at upstream infinity, and $\phi = \phi_o$ sets the azimuthal direction of the undisturbed field at infinity. Note that $\theta_{BV}$ is not a heliocentric angle, and is distinct from the polar coordinate $\theta$.

The solution for the longitudinal magnetic field is straightforward, since a frozen-in field initially parallel to the flow will remain parallel to the flow. Using equation (1) along with $\nabla \cdot B_\ell = 0$, we have

$$\mathbf{B}_\ell = B_o \cos\theta_{BV}\left[\left(\frac{1}{r^2} - \cos\theta\right)\hat{\mathbf{r}} + \sin\theta\, \hat{\boldsymbol{\theta}}\right]. \tag{6}$$

We obtain the transverse field by the method of Euler potentials, whereby any three-dimensional magnetic field can be described by the expression $\mathbf{B} = \nabla\alpha \times \nabla\beta$. Here, $\alpha$ and $\beta$ are scalar functions of position, such that $\alpha$ = constant and $\beta$ = constant describe independent sets of surfaces which are tangent to the field. The exact expressions for these surfaces are somewhat arbitrary, and we will be guided by the geometry of the flow field in our selection of functions. First, we note that two fluid parcels connected by a field line at upstream infinity will continue to be connected by that field line if the parcels remain on their respective streamlines. Thus, one set of surfaces can be composed of those flow streamlines which are connected by the field at upstream infinity. Using the stream function (2) leads to

$$\alpha = \sqrt{1 + 2\frac{\cos\theta - 1}{r^2 \sin^2\theta}}\, r\sin\theta \sin(\phi-\phi_o), \tag{7}$$

where the square root is applied to obtain uniform spacing of the surfaces at infinity. At upstream infinity, $\alpha = p \sin(\phi - \phi_o)$ and these surfaces take the form of stacked parallel planes, each containing $\mathbf{B}_{to}$ and $\mathbf{V}$ (and so equivalent to the $BV$ plane at infinity). These planar surfaces become distorted as they approach the heliopause.

The other Euler potential is constructed to be independent of the $\alpha$-surfaces. To this end, we consider surfaces that at upstream infinity are given by $\beta \rightarrow r \cos\theta =$ constant, defining a set of vertical sheets perpendicular to the flow. If we can mathematically determine the distortion of these sheets as they are carried by the flow past the obstacle, we will have the solution to the problem. In this regard, we define $\beta$ as a "time-like" variable for this steady-state problem, such that $dr/d\beta = V_r$, the radial component of $\mathbf{V}$. It follows that, for fluid parcels starting at upstream infinity,

$$\beta = \int_\infty^r \frac{dy}{V_r(y)}, \tag{8}$$

where the integral is taken along fluid streamlines. Here, the plane defined by $\beta = 0$ connects all the fluid parcels at upstream infinity, so it is parallel to the transverse magnetic field there. Since the integral in equation (8) follows the fluid, setting $\beta(r, \theta)$ to any other constant will map out a plane in space which continues to connect those fluid parcels, and so will remain parallel to the field. Thus, $\beta$ as given by equation (8) will serve as the second Euler potential.[†]

Specifically, writing $V_r$ for constant $\Psi$ we have, using equations (1) and (2),

$$\beta(r,\theta) = \int_\infty^r \frac{y^2 dy}{(\mp)\sqrt{y^4 + 2(\Psi-1)y^2 + 1}} = \int_\infty^r \frac{y^2 dy}{(\mp)\sqrt{y^4 - (r^2 \sin^2\theta + 2\cos\theta)y^2 + 1}}. \tag{9}$$

We see that each streamline contains a point, at the surface $r^2 \cos\theta - 1 \equiv \Delta = 0$, where $V_r$ passes through zero as it moves past the heliosphere. The $(\mp)$ in equation (9) indicates that, following the behavior of $V_r$, the negative square root should be taken upstream of that point, continuing the integral with the positive square root when $(r, \theta)$ is past that point. Nothing unusual happens to the flow or the magnetic field across this surface, as will be seen below. The singular behavior of the potential is an artifact of our definition (8), and does not lead to any discontinuity in the flow or field. A similar feature appears at $\theta = \pi/2$ in the solution of the spherical obstacle problem (see Appendix).

---

[†] Concerning the limits of the integral in equation (8), we note that the indefinite integral gives $\beta = r \cos\theta$ for large $r$ as stated. By setting the lower limit to $\infty$, we are formally subtracting an infinite constant from $\beta$. However, since the field is unaffected by the addition of arbitrary constants to the Euler potentials (in the same sense that potentials in other applications are only defined to within an arbitrary additive constant), we use the convenient expression above. This step will also lead to the improved expression for the case of the spherical obstacle given in the Appendix.

The solution for the transverse field is then given by $\mathbf{B}_t = \nabla\alpha \times \nabla\beta$. When normalized to the value of the transverse field at upstream infinity, the solution upstream of the $\Delta = 0$ surface takes the form

$$\frac{\mathbf{B}_t}{B_o \sin\theta_{BV}} = -\left(1+2\frac{\cos\theta-1}{r^2\sin^2\theta}\right)^{1/2}\cos(\phi-\phi_o)\left\{r\sin\theta\left(\cos\theta-r^{-2}\right)g(r,\theta)\hat{\mathbf{r}}\right.$$

$$\left.-\left[\left(\cos\theta-\frac{1}{r^2}\right)^{-1}+r\sin^2\theta\, g(r,\theta)\right]\hat{\boldsymbol{\theta}}\right\} - \left(1+2\frac{\cos\theta-1}{r^2\sin^2\theta}\right)^{-1/2}\sin(\phi-\phi_o)\hat{\boldsymbol{\phi}}, \qquad (10)$$

where

$$g(r,\theta) = \int_\infty^r \frac{y^4\, dy}{\left[y^4 - \left(r^2\sin^2\theta + 2\cos\theta\right)y^2 + 1\right]^{3/2}}, \qquad (11)$$

As $r \to \infty$, we see that $g(r, \theta) \to -(r\cos\theta)^{-1}$, so the expression (10) satisfies the boundary condition (5) for the upstream magnetic field.

The lower limit of the integral $g$ refers to upstream infinity, where the flow originates and the streamlines begin. When $r^2\cos\theta - 1 = \Delta < 0$, one needs to propagate this solution through the point $V_r = 0$. Care must also be taken when $\Delta$ is positive but small. Therefore, we define three regions such that $\Delta > \varepsilon$, $|\Delta| < \varepsilon$, and $\Delta < -\varepsilon$, where $\varepsilon$ is a small positive number (taken to be $\varepsilon = 0.005$ in the examples to follow). The expressions (10) and (11) give the solution when $\Delta > \varepsilon$.

In the vicinity of the $V_r = 0$ surface, $|\Delta| < \varepsilon$, and we expand these expressions for small $\Delta$. Rewriting the functions of $\theta$ in terms of $\Delta$, we see that the denominator in the integrand of $g$ can be factored into two terms, each quadratic in $y$, only one of which approaches zero when $y \to r$ at $\Delta = 0$. Specifically, these factors are

$$y^2 - \frac{r^4-1}{2r^2}\left(\frac{r^4+1-\Delta^2}{r^4-1} \pm \sqrt{1 - 2\frac{r^4+1}{(r^4-1)^2}\Delta^2 + \frac{\Delta^4}{(r^4-1)^2}}\right). \quad (12)$$

Thus, the denominator becomes

$$\left[y^2 - r^2 + \frac{r^2}{r^4-1}\Delta^2 + \frac{r^2}{(r^4-1)^3}\Delta^4\right]^{3/2} \left[y^2 - \frac{1}{r^2} - \frac{1}{r^4-1}\frac{\Delta^2}{r^2} - \frac{r^2}{(r^4-1)^3}\Delta^4\right]^{3/2} \quad (13)$$

for small $\Delta$, where $r$ and $\Delta$ are constant during the integration. We then integrate by parts twice to remove the first factor from the denominator, so the remaining integral is well behaved. This procedure yields, to second order in $\Delta$,

$$\left(\cos\theta - r^{-2}\right)g(r,\theta) = -\frac{r^3}{r^4-1} - \frac{4r^3\Delta^2}{(r^4-1)^3} + \frac{3\Delta}{r^4}h(r,r), \quad (14)$$

and

$$\left[\left(\cos\theta - \frac{1}{r^2}\right)^{-1} + r\sin^2\theta\, g(r,\theta)\right] = \frac{r^2}{r^4-1}\left[2 + \frac{(r^4-5)\Delta}{r^4-1} + \frac{8\Delta^2}{(r^4-1)^2}\right]$$

$$+ 3\frac{r^4-1-2\Delta}{r^5}h(r,r), \quad (15)$$

where

$$h(r,d) = \int_\infty^d \frac{\left[4y^2 + \frac{1}{r^2}\left(1 + \frac{\Delta^2}{r^4-1}\right)\right]\sqrt{\left|y^2 - r^2\left(1 - \frac{\Delta^2}{r^4-1} + \frac{\Delta^4}{(r^4-1)^3}\right)\right|}}{(\mp)\left[y^2 - \frac{1}{r^2}\left(1 + \frac{\Delta^2}{r^4-1}\right)\right]^{7/2}} dy. \quad (16)$$

The $(\mp)$ expression within the integral (16) is understood in the same manner as in equation (9): When $\Delta > 0$, the minus sign is used. When $\Delta < 0$, the integral is taken with the minus sign to the point $r = r_o$, corresponding to $\Delta = 0$, and then continued with the plus sign from that point to the upper limit. A given streamline, labeled by $\Psi$ in equation (2), encounters the $\Delta = 0$ point at

$$r_o^2 = (1-\Psi) + \sqrt{(1-\Psi)^2 - 1}. \tag{17}$$

These expressions should be plugged into the corresponding factors in the components of equation (10) for values of $r$ and $\theta$ near enough to the $\Delta = 0$ surface. It is clear from these expressions that the transverse field (10) is finite and continuous around $\Delta = 0$.

In the region downstream of the $V_r = 0$ surface, $\Delta < -\varepsilon$, we define a quantity $r_d$ which is the radial position of the streamline leading to the point in question, when that streamline crosses the surface at $\Delta = -\varepsilon$. With the same procedure leading to equation (17), this radius is given by

$$r_d^2 = (1-\Psi) + \sqrt{(1-\Psi)^2 + \varepsilon^2 - 1}. \tag{18}$$

We then split the integral $g(r, \theta)$ in equation (10) into the piece from upstream infinity to $r_d$ (already downstream of the $\Delta = 0$ surface) and the piece from $r_d$ to $r$. We write the first piece of the integral using the expansions (14) and (15) evaluated at $r = r_d$. The downstream solution is then given by equation (10) with the function $g(r, \theta)$ replaced by

$$g(r,\theta) = \frac{r_d^5}{\varepsilon(r_d^4 - 1)} + 4\frac{r_d^5 \varepsilon}{(r_d^4 - 1)^3} - \frac{3}{r_d^2}\left(1 + \frac{\varepsilon^2}{r^4 - 1}\right) h(r,r_d) - g^*(r,\theta) \tag{19}$$

where $g^*$ is the integral in equation (11) with the lower limit set to $r_d$. The minus sign in front of $g^*$ is a consequence of our original definition of $g$ in equation (11) as specific to the upstream region.

In Figure 2, we show some sample field lines draping over the heliopause for an orientation of the undisturbed ISMF as indicated by the center of the IBEX ribbon. The heliocentric spherical coordinate system here has $\theta = 0$ pointing from the Sun to the nose of the heliosphere (assumed given by the observed interstellar helium velocity) and the $\phi = 0$ direction pointing to the port side of the heliosphere parallel to the heliographic equator from the nose (that is, in the direction of increasing ecliptic longitude). In this system, we set $\theta_{BV} = 125°$ and $\phi_o = 153°$, choosing this field polarity to be consistent with the Voyager 1 observations discussed in Section 5.

## 4. Implications for the IBEX ribbon

12Despite the many idealizations in this model, particularly our neglect of magnetic pressure effects, it is still possible to obtain useful information. We will see that several qualitative properties can be reproduced here, indicating that the overall structure of our model field is related to real magnetic field around the heliosphere.

In general, we find that the draping effects in our model are concentrated much closer to the heliopause than indicated by observations of the real ISMF. This concentration is, of course, expected due to the lack of a backreaction of the magnetic pressure on the fluid flow. In the real LISM, magnetic forces will cause the flow to slow and turn much sooner than found in this passive field case. Thus, the real deflection will start further upstream and take place more gradually than in our model, and the observable changes will be larger.

The first comparison we make is with the geometry of the IBEX ribbon as determined by the analysis of Funsten et al. (2013). In that work, the positions of the peak ribbon flux in many intervals along the ribbon were plotted and found to a good approximation to map out circles in the plane of the sky. The opening angle of the circles increased with increasing energy of the detected ENAs, from ~ 73° in the energy range 0.7 - 1.7 keV to almost 80° in the highest energy bin at 4.3 keV. The positions of the circle centers were also energy dependent, drifting in a direction roughly transverse to the heliospheric nose position by about 10° from the lowest to highest energy. Within the secondary ENA scenario for the ribbon, the peak flux positions should indicate where the local ISMF is perpendicular to the heliocentric radial direction. If the magnetic field at the source of the ribbon particles were not affected by draping around the heliopause, the peak flux positions would be arranged in a great circle (opening angle = 90°) according to this scenario. The observed opening angles less than 90° then indicate the effects of magnetic field draping at the source of the ENAs. Further, an opening angle which is closer to 90° for the more energetic particles is consistent with the interpretation that these particles were emitted at greater distances from the Sun where the draping effects are smaller.

In Figure 3, we plot the position of $\mathbf{B} \cdot \hat{\mathbf{r}} = 0$ from our model for four distances from the Sun, using the upstream orientation of the ISMF from the high-energy ribbon center, $\theta_{BV} = 125°$ and $\phi_o = 153°$, as in Figure 2. The left-hand panel shows these curves in the nose-centered coordinate system, and the right-hand panel shows them in a system rotated by $(\theta, \phi) = (55°, -27°)$. We see that the locus of perpendicular field in our draped field model gives curves which are nearly circles on the sky. The innermost curve is not closed in the downwind direction, since the fluid deflection by the tail of the model heliopause prevents the field at that distance and direction from turning perpendicular to $\hat{\mathbf{r}}$. There may be a similar effect in the Funsten et al. (2013) observations, though possible ribbon emission in this direction is also obscured by the Earth's magnetosphere.

The circles in Figure 3 are clearly smaller for closer positions, though this effect is less pronounced than the behavior observed by Funsten et al. (2013). The model circle at $r = 10$ has an opening angle of almost 90°. The closer circles have approximate opening angles of 88°, 87°, and 85°, respectively. Although this trend continues for closer positions, the circular segments become shorter for $r < 3$ so we do not show these results.

A drift of the circle centers for different distances, possibly corresponding to different ENA energies, is also evident in Figure 3. However, in this model the drift appears in the direction of the heliospheric nose, rather than transverse to that direction as observed by Funsten et al. (2013). The cause of this discrepancy is not clear, but it may be due to the simple shape of our model heliopause. The real heliopause is not believed to be axisymmetric with respect to the interstellar flow, and we suspect that a model flow around a non-axisymmetric obstacle may yield a field structure with behavior closer to that observed. We will report on magnetic field draping around non-axisymmetric obstacles in a future publication.

## 5. Implications for the Voyager measurements

The two Voyager spacecraft continue to travel away from the Sun in the general direction of the heliospheric nose. As of this writing, Voyager 1 has apparently crossed the heliopause into interstellar space, while Voyager 2 remains in the inner heliosheath region between the solar wind termination shock and the heliopause. The magnetometers on both spacecraft are still operating and are providing a fascinating picture of the *in situ* magnetic field along their trajectories.

Without a functioning plasma instrument to provide corroboration, the passage of Voyager 1 through the heliopause was uncertain, and remains somewhat controversial. A particularly puzzling aspect was that the magnetic field angle did not change noticeably, although the field intensity did increase substantially (Burlaga et al. 2013). Prior to the crossing, the measured heliosheath field direction was fairly close to the nominal Parker spiral expected for that heliolatitude, and the fact that it did not rotate toward the undisturbed interstellar field direction in an appreciable manner led to a number of hypotheses as to what had happened (Schwadron & McComas 2013b; Swisdak et al. 2013; Strumik et al. 2013; Strumik et al. 2014; Opher & Drake 2013; Borovikov & Pogorelov 2014; Grygorczuk et al. 2014; Fisk & Gloeckler 2014).

It is generally understood that the interstellar field beyond the heliopause would be modified due to draping effects, and an obvious question is whether the draping could explain the Voyager data. By its passive nature, the model presented here cannot treat the heliopause transition itself or reliably discuss details in the immediate vicinity of that surface. Our model heliopause is in fact a singular surface in that the field magnitude increases without bound in the absence of the backreaction on the imposed flow. However, the geometrical trend of the field angles as the field lines bend around the heliosphere can give a qualitative prediction of the behavior of the local field as one moves along the path of the two Voyagers.

In Figure 4, we show the behavior of the model interstellar field along radial lines in the directions of the Voyager 1 (Fig. 4a) and Voyager 2 (Fig. 4b) trajectories. In our nose-oriented coordinate system we take the Voyager 1 direction as $(\theta, \phi) = (30°, 82°)$ and the Voyager 2 direction as $(51°, -52°)$. The model heliopause is found at $r = 1.035$ in the Voyager 1 direction, and at $r = 1.108$ in the Voyager 2 direction, according to equation (3). The field direction in the Figure is given in terms of $\lambda$ and $\delta$, the local spacecraft frame azimuthal and elevation angles, respectively, as defined by Burlaga (1995).

As stated, the model interstellar field magnitude increases abruptly for radii just beyond the heliopause. At the Voyager 1 position (Figure 4a), we see that the draped magnetic field in the model starts with an azimuthal angle near the Parker spiral direction, and an elevation angle somewhat above the ideal Parker value. Both angles change smoothly with distance to approach the direction of the undisturbed ISMF on that radial line. We have also indicated in red the angles measured at Voyager 1 for the last half of 2013 $(\lambda, \delta) = (292°, 22°)$, as reported by Burlaga & Ness (2014b).

It is interesting that the observed field angles are reasonably close to those predicted by the model. The observed azimuthal angle is squarely in the range suggested for the draped field, and the observed elevation angle is only high by ~ 10° or less. We note that the elevation angle at Voyager 1 was consistently higher than the ideal Parker value both before and after the presumptive heliopause transition. Perhaps more significant for a qualitative comparison is the fact that the observed angles trend in the same direction as those predicted by the model, with $\lambda$ increasing and $\delta$ decreasing as Voyager 1 moves further from the Sun (Burlaga & Ness 2014b).

For the case of Voyager 2, our model presents the opportunity to make a prediction in advance of the observations. Figure 4b shows the model field structure in the direction of Voyager 2. We see that the predicted azimuthal and elevation angles of the interstellar field quickly take on the values of the distant field and hardly change at all as the radial position increases. This behavior is due to the fact that the line connecting the Voyager 2 path with the nose of the model heliosphere is nearly parallel to the undisturbed ISMF. In this case, the draped field along that path tends to remain in a single plane - it will be stretched but not substantially twisted by the flow around the obstacle. We do not know what the heliosheath field may be like as Voyager 2 approaches the heliopause, so we cannot say whether those data will yield a clearer signature of the heliopause crossing than found at Voyager 1. However, this result suggests that the field angles measured at Voyager 2 beyond the heliopause will not be much affected by draping, and will point in the direction of the undisturbed ISMF almost immediately after crossing the boundary.

Of course, this claim depends on our model assumptions: primarily that the actual heliopause is reasonably axisymmetric, and that the IBEX ribbon indicates the direction of the ISMF. Our prediction may need to be modified for cases of non-axisymmetric obstacles, and we will explore such non-axisymmetric effects in future work.

## 5. Summary and conclusions

We have described a model for the draping of the interstellar magnetic field around the heliosphere, under the assumptions that the heliopause is axisymmetric in the interstellar flow, that the flow around the heliopause is ideal, and that the magnetic field frozen into that flow exerts no substantial stresses on the system. We then obtained the analytical solution for this field in the case where the heliopause shape is a Rankine half-body.

Setting the orientation of the model ISMF at infinity along the direction indicated by the IBEX ribbon, we compared the qualitative structure of the draped field to several observations by the IBEX and Voyager spacecraft. We found that the places where the model draped field at constant distance from the Sun was perpendicular to the heliocentric radial direction, $\mathbf{B} \cdot \hat{\mathbf{r}} = 0$, marked out nearly circular curves on the plane of the sky. These circles were progressively smaller for closer positions. Interpretation of the apparent shapes of the IBEX ribbon in the context of the secondary ENA scenario for the ribbon suggests that the actual draped ISMF has this character (Funsten et al. 2013). However, the concurrent progression of the circle centers across the sky did not follow the observed behavior, which we suspect was due to the lack of asymmetry in our present model. These results provide support both for our qualitative ISMF model and for the secondary ENA hypothesis of the IBEX ribbon.

The qualitative comparison of our model with the magnetic field measurements at Voyager 1 was encouraging, and indicates that the initially unexpected behavior of the observed field could perhaps be explained simply by the effect of field-line draping. These results strengthen the interpretation that the spacecraft has actually crossed the heliopause (Burlaga & Ness 2014b).

The heliopause in our model was a smooth, continuous surface and did not include any small-scale distortions, such as might result from magnetic reconnection (Swisdak et al. 2013) or MHD instabilities (Borovikov & Pogorelov 2014). These small-scale processes could certainly still be present at the real heliopause. However, we emphasize that the draping effects in our model do not depend in any way on a "frictional" interaction or other large-scale influence of the solar magnetic field on the ISMF, as has been suggested by Opher & Drake (2013). It appears that the Voyager 1 observations may be understood without invoking such additional effects.

Finally, we tentatively predict on the basis of this model that the angular effect of draping along the path of Voyager 2 will be small. Thus, we expect that the magnetic field encountered by Voyager 2 very soon after crossing the heliopause will be substantially aligned with the direction of the undisturbed ISMF.

**APPENDIX**

**An improved expression for the passive magnetic field around a spherical obstacle**

The solution for a passive magnetic field frozen into the potential flow around a sphere was (to the best of our knowledge) first worked out by Bernikov & Semenov (1979) and has been reiterated in several works since then (Chacko & Hassam 1997; Dursi & Pfrommer 2008; Romanelli et al. 2014). These works assumed that the plasma flow and the magnetic field were uniform and constant at upstream infinity. In the spherical coordinate system with the origin at the center of a unit sphere, and the polar axis pointing into the flow, the normalized potential flow for this case is given by

$$\mathbf{V} = -\left(1 - \frac{1}{r^3}\right)\cos\theta\, \hat{\mathbf{r}} + \left(1 + \frac{1}{2r^3}\right)\sin\theta\, \hat{\boldsymbol{\theta}}, \tag{A1}$$

and the stream function by

$$\Psi = -\left(1 - \frac{1}{r^3}\right)\frac{r^2 \sin^2\theta}{2}. \tag{A2}$$

The first steps in this calculation are completely analogous to those of our Rankine problem. The magnetic field can be decomposed into longitudinal and transverse components, and the longitudinal solution is trivially obtained in the same way as above. The transverse solution follows through an integration of the coupled differential equations along the fluid characteristics, and may be written (after Dursi & Pfrommer (2008)) as

$$\frac{B_r}{B_o} = \left(1 - \frac{1}{r^3}\right)\cos\theta \left[C_1 \mp p\sin\phi \int_\xi^r \frac{y^4 dy}{\sqrt{y^3 - 1}\left(y^3 - p^2 y - 1\right)^{3/2}}\right] \tag{A3}$$

$$\frac{B_\theta}{B_o} = \frac{2r^3 + 1}{r^{5/2}\sqrt{r^3 - 1}}\left[C_2 \pm 2\sin\phi \int_\xi^r \frac{y^3(y^3 + 2)\sqrt{y^3 - 1}\, dy}{\left(2y^3 + 1\right)^2 \left(y^3 - p^2 y - 1\right)^{1/2}}\right] \tag{A4}$$

$$\frac{B_\phi}{B_o} = \frac{\cos\phi}{\sqrt{1 - 1/r^3}} \tag{A5}$$

where the equivalent impact parameter $p(r,\theta) = \sqrt{-2\Psi} = r\sin\theta\,(1 - r^{-3})$ is a constant along streamlines, $C_{1,2}$ are integration constants to be determined by the upstream boundary condition and $\xi$ is the value of $r$ at the upstream boundary.

As in the Rankine problem, one must be careful in the region where $V_r$ passes through zero, which happens here at $\theta = \pi/2$. The previous works state that the upper sign in equations (A3) and (A4) refers to the region upstream of the plane at $\theta = \pi/2$, and the lower sign to the region downstream of that plane. Since the value of $\xi$ is not rigorously specified here, and the integrals may be applied in a piecewise manner, the above expressions are formally correct.

However, the notation here is somewhat misleading and it would be more appropriate to use some form of our $(\mp)$ notation inside the integrals, as defined under equation (9). In that way, the upstream boundary at $r = \xi$ can be set at a constant position and the integrals propagated unambiguously to any downstream point.

The above solution has another problem, though, in that the integral in equation (A4) diverges as $\xi \to \infty$. Thus, even though it should be perfectly reasonable to set the upstream boundary at infinity, this expression is not formally defined there. This makes it difficult to apply the upstream boundary condition for $B_\theta$, and the previous workers resorted to expansions close to the spherical obstacle to evaluate it. (The boundary condition on $B_r$ is straightforward, giving $C_1 = 0$.) This difficulty can be eliminated through the procedure in this paper, using the Euler potentials. By starting with potential functions, one can rigorously subtract the infinite constant which creates the divergent result, as we did in equation (8).

Following the procedure of Section 3, we define the Euler potentials for the spherical case as

$$\alpha = r \sin\theta \cos\phi \sqrt{1 - \frac{1}{r^3}} \tag{A6}$$

$$\beta = \int_\infty^r \frac{y^3 dy}{(\mp)\sqrt{(y^3-1)(y^3 - p^2 y - 1)}}, \tag{A7}$$

where here we take the orientation of the field at infinity to be along $\phi = \pi/2$ for consistency with Dursi & Pfrommer (2008). Now, the solution for the transverse field component, given by $\mathbf{B} = \nabla\alpha \times \nabla\beta$, results in the well-behaved expressions

$$\frac{B_r}{B_o} = \left(1 - \frac{1}{r^3}\right) p \cos\theta \sin\phi \int_\infty^r \frac{y^4 dy}{(\mp)\sqrt{y^3 - 1}\left(y^3 - p^2 y - 1\right)^{3/2}} \tag{A8}$$

$$\frac{B_\theta}{B_o} = \left[\frac{1}{\cos\theta \sqrt{1 - 1/r^3}} + \left(1 + \frac{1}{2r^3}\right) p \sin\theta \int_\infty^r \frac{y^4 dy}{(\pm)\sqrt{y^3 - 1}\left(y^3 - p^2 y - 1\right)^{3/2}}\right] \sin\phi \tag{A9}$$

to replace equations (A3) and (A4), where (A5) remains the same.

**Acknowledgements.** We are grateful for valuable conversations with L. F. Burlaga, S. A. Fusilier, J. R. Jokipii, J. Kleimann, H. Kucharek, M. A. Lee, and N. A. Schwadron. This work was initiated in conversations at the 13th Annual International Astrophysics Conference in Myrtle Beach, SC during March 2014, and we thank Gary Zank and Adele Corona for organizing that conference. This work was supported in part by NASA grant NNX13AF97G, NSF grant AGS1358103, and by the IBEX mission through NASA contract NNG05EC85C.

# References

Batchelor, G. K. 1967, An Introduction to Fluid Mechanics (Cambridge: Cambridge Univ. Press)
Ben-Jaffel, L., Strumik, M., Ratkiewicz, R., & Grygorczuk, J. 2013, Astrophys. J., 779, 130
Bernikov, L. V., & Semenov, V. S. 1979, Geomagn. Aeron., 19, 452
Borovikov, S., & Pogorelov, N. V. 2014, Astrophys. J., 783, L16
Burlaga, L. F. 1995, Interplanetary Magnetohydrodynamics (New York: Oxford Univ. Press)
Burlaga, L. F., & Ness, N. F. 2014a, Astrophys. J., 784, 146
———. 2014b, Astrophys. J., 795, L19
Burlaga, L. F., Ness, N. F., & Stone, E. C. 2013, Science, 341, 147
Bzowski, M., et al. 2012, Astrophys. J. Suppl., 198, 12
Chacko, Z., & Hassam, A. B. 1997, Phys. Plasmas, 4, 3031
Chalov, S. V., Alexashov, D. B., McComas, D. J., Izmodenov, V., Malama, Y. G., & Schwadron, N. A. 2010, Astrophys. J., 716, L99
Dursi, L. J., & Pfrommer, C. 2008, Astrophys. J., 677, 993
Fisk, L. A., & Gloeckler, G. 2014, Astrophys. J., 789, 41
Florinski, V., Zank, G. P., Heerikhuisen, J., Hu, Q., & Khazanov, I. 2010, Astrophys. J., 719, 1097
Frisch, P. C., et al. 2013, Science, 341, 1080
———. 2015, Astrophys. J., 801, 61
Funsten, H. O., et al. 2009, Science, 326, 964
———. 2013, Astrophys. J., 776, 30
Fuselier, S. A., et al. 2009, Science, 326, 962
Gamayunov, K., Zhang, M., & Rassoul, H. 2010, Astrophys. J., 725, 2251
Gloeckler, G., et al. 2004, Astron. Astrophys., 426, 845
Grygorczuk, J., Czechowski, A., & Grzedzielski, S. 2014, Astrophys. J., 789, L43
Grygorczuk, J., Ratkiewicz, R., Strumik, M., & Grzedzielski, S. 2011, Astrophys. J., 727, L48
Heerikhuisen, J., & Pogorelov, N. V. 2011, Astrophys. J., 738, 29
Heerikhuisen, J., Zirnstein, E. J., Funsten, H. O., Pogorelov, N. V., & Zank, G. P. 2014, Astrophys. J., 784, 73
Heerikhuisen, J., et al. 2010, Astrophys. J., 708, L126
Isenberg, P. A. 2014, Astrophys. J., 787, 76
Izmodenov, V. V., Alexashov, D. B., & Myasnikov, A. 2005, Astron. Astrophys., 437, L35
Lallement, R., & Bertaux, J. L. 2014, Astron. Astrophys., 565, A41
Lallement, R., Quémerais, E., Bertaux, J. L., Ferron, S., Koutroumpa, D., & Pellinen, R. 2005,

Science, 307, 1447
Lallement, R., Quémerais, E., Koutroumpa, D., Bertaux, J. L., Ferron, S., Schmidt, W., & Lamy, P. 2010, in Solar Wind 12, ed. M. Maksimovic, K. Issautier, N. Meyer-Vernet, M. Moncuquet, & F. Pantellini (Melville, NY: Amer. Instit. Phys.), 555
Leonard, T., et al. 2015, Astrophys. J., in press
McComas, D. J., Lewis, W. S., & Schwadron, N. A. 2014, Rev. Geophys., 52, 118
McComas, D. J., et al. 2009, Science, 326, 959
———. 2015, Astrophys. J., 801, 28
Mitchell, J. J., Cairns, I. H., Pogorelov, N. V., & Zank, G. P. 2008, J. Geophys. Res., 113, A04102
Möbius, E., Liu, K., Funsten, H. O., Gary, S. P., & Winske, D. 2013, Astrophys. J., 766, 129
Möbius, E., et al. 2012, Astrophys. J. Suppl., 198, 11
———. 2004, Astron. Astrophys., 426, 897
Opher, M., & Drake, J. F. 2013, Astrophys. J., 778, L26
Opher, M., Stone, E. C., & Liewer, P. C. 2006, Astrophys. J., 640, L71
Opher, M., Stone, E. C., & Gombosi, T. I. 2007, Science, 316, 875
Opher, M., Bibi, F. A., Toth, G., Richardson, J. D., Izmodenov, V. V., & Gombosi, T. I. 2009, Nature, 462, 1036
Parker, E. N. 1961, Astrophys. J., 134, 20
Pogorelov, N. V., Heerikhuisen, J., & Zank, G. P. 2008, Astrophys. J., 675, L41
Pogorelov, N. V., Stone, E. C., Florinski, V., & Zank, G. P. 2007, Astrophys. J., 668, 611
Pogorelov, N. V., Heerikhuisen, J., Mitchell, J. J., Cairns, I. H., & Zank, G. P. 2009, Astrophys. J., 695, L31
Pogorelov, N. V., Heerikhuisen, J., Zank, G. P., Borovikov, S., Frisch, P. C., & McComas, D. J. 2011, Astrophys. J., 742, 104
Ratkiewicz, R., & Grygorczuk, J. 2008, Geophys. Res. Lett., 35, L23105
Ratkiewicz, R., Strumik, M., & Grygorczuk, J. 2012, Astrophys. J., 756, 3
Röken, C., Kleimann, J., & Fichtner, H. 2015, Astrophys. J., in press
Romanelli, N., Gómez, D., Bertucci, C., & Delva, M. 2014, Astrophys. J., 789, 43
Schwadron, N. A., & McComas, D. J. 2013a, Astrophys. J., 764, 92
———. 2013b, Astrophys. J., 778, L33
Schwadron, N. A., et al. 2014, Science, 343, 988
———. 2009, Science, 326, 966
Stern, D. P. 1966, Space Sci. Rev., 6, 147
———. 1970, Amer. J. Phys., 38, 494
Strumik, M., Ben-Jaffel, L., Ratkiewicz, R., & Grygorczuk, J. 2011, Astrophys. J., 741, L6
Strumik, M., Czechowski, A., Grzedzielski, S., Macek, W. M., & Ratkiewicz, R. 2013, Astrophys. J., 773, L23
Strumik, M., Grzedzielski, S., Czechowski, A., Macek, W. M., & Ratkiewicz, R. 2014, Astrophys. J., 782, L7
Swisdak, M., Drake, J. F., & Opher, M. 2013, Astrophys. J., 774, L8
Whang, Y. C. 2010, Astrophys. J., 710, 936
Witte, M. 2004, Astron. Astrophys., 426, 835
Wood, B. E., Müller, H. R., & Witte, M. 2015, Astrophys. J., 801, 62
Zhang, M., Zuo, P., & Pogorelov, N. V. 2014, Astrophys. J., 790, 5

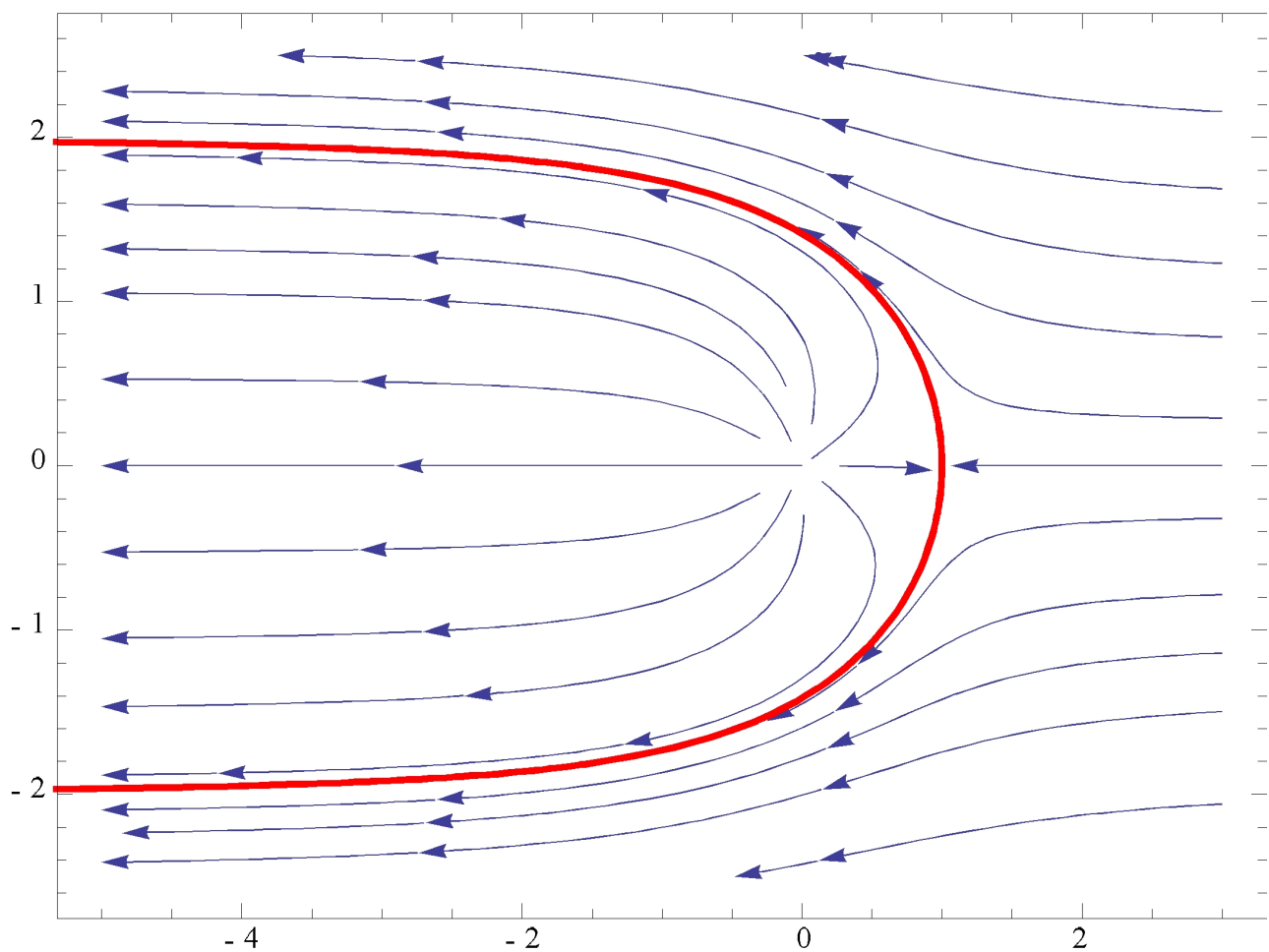

**Figure 1.** Illustration of the potential fluid flow around a Rankine half-body. The surface of the model heliopause is plotted in red.

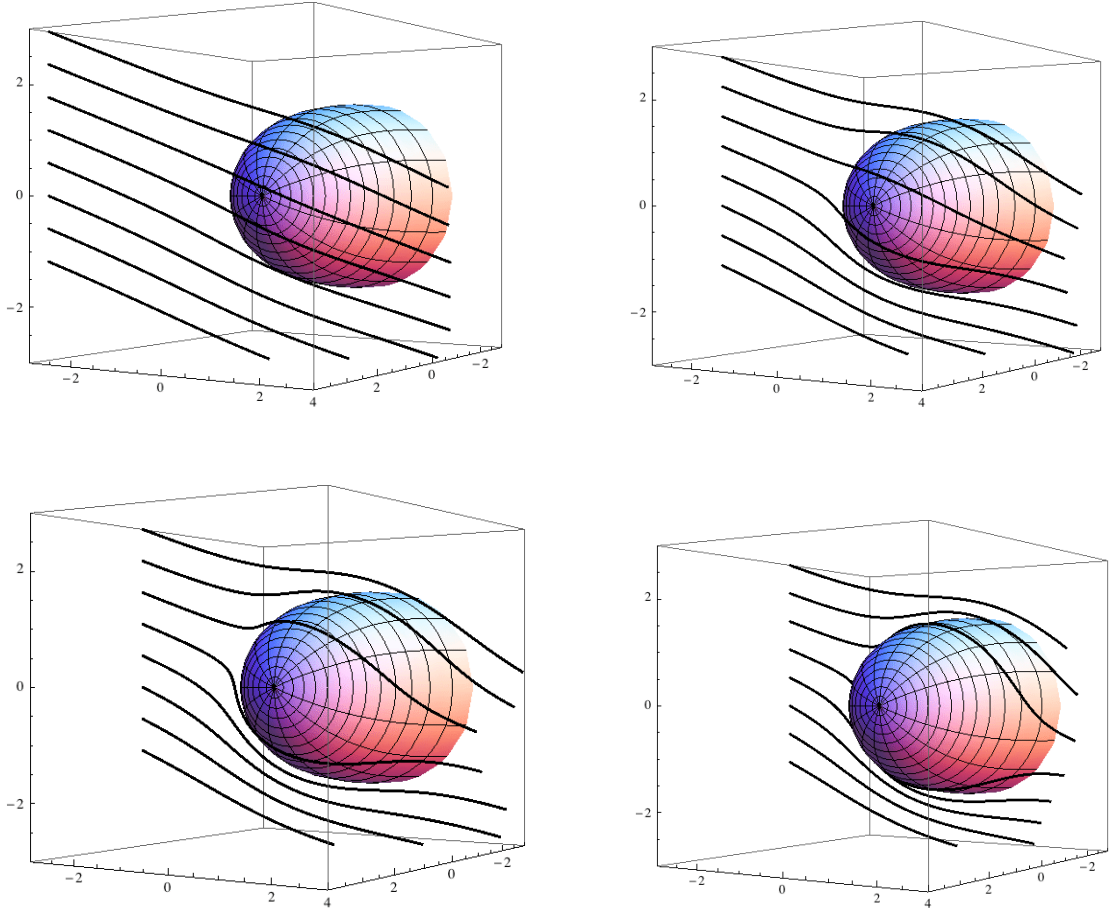

**Figure 2.** Sample field lines as the ISMF is carried around the model heliopause, as viewed from a position upstream, in the heliocentric equator and off the port side of the heliosphere. The undisturbed field at infinity is set at $\theta_{BV} = 125°$ to the inflow direction and $\phi_o = 153°$ to the heliocentric equatorial plane.

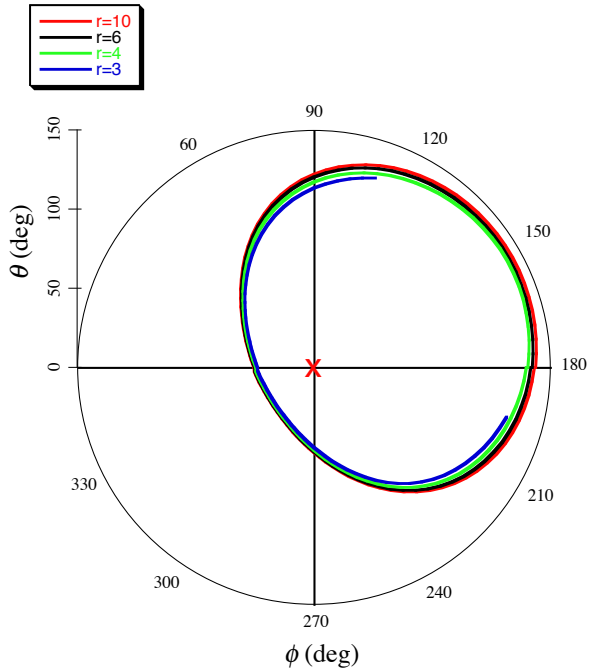 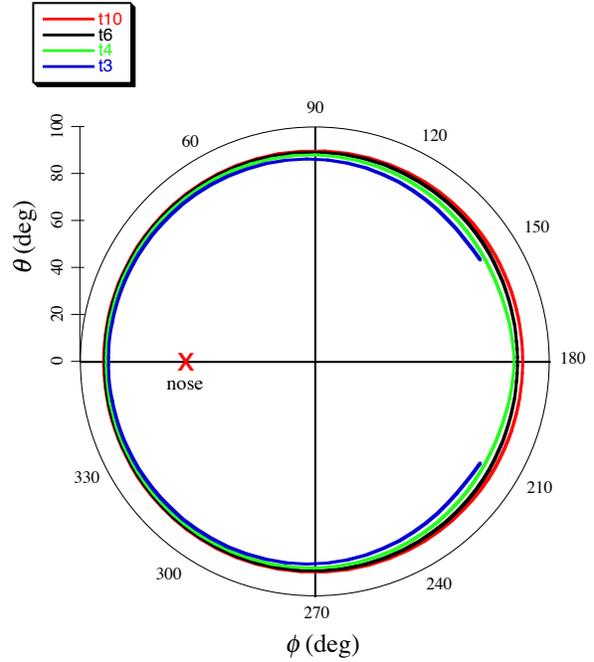

**Fig. 3a**  **Fig. 3b**

**Figure 3.** Locations of $\mathbf{B} \cdot \hat{\mathbf{r}} = 0$ on the plane of the sky as viewed from the Sun for four heliocentric distances: $r = 10$ (red), $r = 6$ (black), $r = 4$ (green), and $r = 3$ (blue), in units of the model distance to the heliospheric nose. The left panel shows the curves in coordinates centered on the nose, and the right panel shows the same curves in a coordinate system rotated by $(\theta, \phi) = (55°, -27°)$. The red × indicates the position on the sky of the nose.

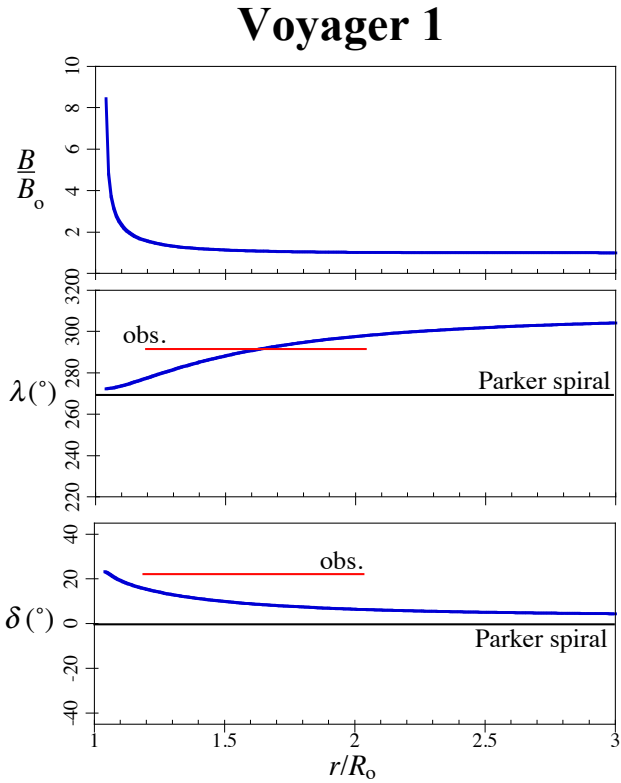 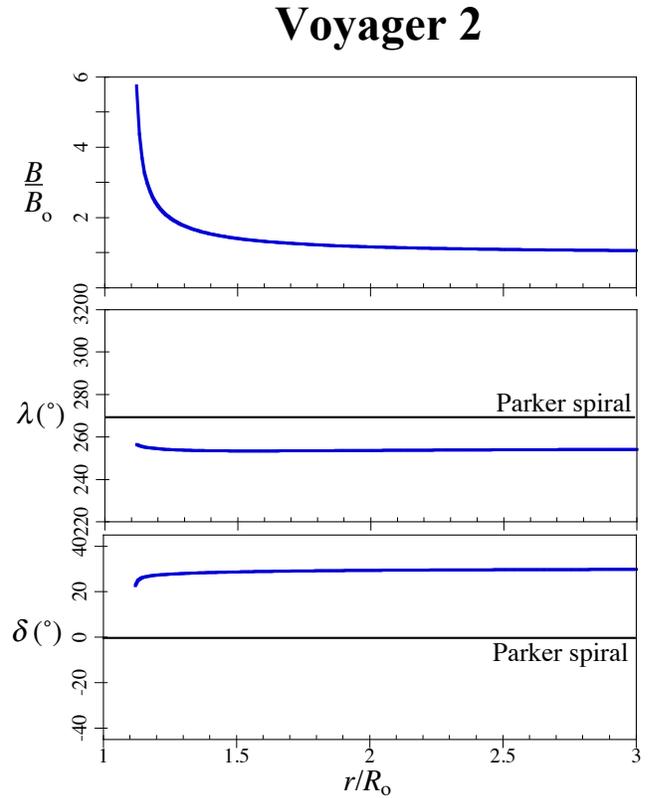

**Fig. 4a**  **Fig. 4b**

**Figure 4.** Model magnetic fields along radial lines in the directions of the two Voyager spacecraft. The angular field directions are in spacecraft *RTN* coordinates. The left panel shows Voyager 1 along the radial line $(\theta, \phi) \sim (30°, 82°)$, and the right panel shows Voyager 2 along the line $(\theta, \phi) \sim (51°, -52°)$. The red lines in the left panel show the observed field as reported by Burlaga & Ness (2014).